\documentclass[aps,prl,twocolumn,showpacs]{revtex4}
\usepackage[dvips]{graphicx}

\begin{document}
\draft
\raggedbottom

\title{Catching proteins in liquid helium droplets}

\author{Frauke Bierau}
\author{Peter Kupser}
\author{Gerard Meijer}
\author{Gert von Helden}
\email{E-mail: helden@fhi-berlin.mpg.de}

\affiliation{Fritz-Haber-Institut der Max-Planck-Gesellschaft, Faradayweg 4-6, D-14195 Berlin, Germany}

\date{\today}

\begin{abstract}
An experimental approach is presented that allows for the incorporation of large mass/charge selected ions in liquid helium droplets. It is demonstrated that droplets can be efficiently doped with a mass/charge selected amino acid as well as with the much bigger m$\approx$12 000 amu protein Cytochrome C in selected charge states. The sizes of the ion-doped droplets are determined via electrostatic deflection. Under the experimental conditions employed, the observed droplet sizes are very large and range, dependent on the incorporated ion, from 10$^{10}$ helium atoms for protonated Phenylalanine  to 10$^{12}$ helium atoms for Cytochrome C. As a possible explanation, a simple model based on the size- and internal energy-dependence of the pickup efficiency is given.
\end{abstract}


\maketitle

The isolation of foreign species in liquid helium nano\-droplets is of fundamental interest and has found important applications in molecular spectroscopy \cite{Rev_Angewandte_Toennies_Vilesov, Rev_Toennies_spec, Rev_Stienkemeier, Rev_Choi, Rev_Barranco}. Helium clusters have an internal equilibrium temperature of $\sim$ 0.37~K, which is maintained by evaporative cooling.
Liquid helium is optically transparent from the deep UV to the far IR and superfluid helium droplets can serve as gentle matrices to provide an isothermal environment for embedded molecules at cryogenic temperatures. Further, due to the weak interactions of liquid helium,  molecules embedded in helium nanodroplets can rotate freely and their optical spectra show narrow linewidths \cite{Scoles_SF6, Vilesov_SF6_PRL}.

Helium droplets can be formed in free-jet expansions \cite{Becker_1961} of high pressure pre-cooled helium gas, and this technique is used by many groups to produce continuous or pulsed droplet beams.
The choice of expansion parameters, in particular the orifice diameter d, source pressure P$_0$ and temperature T$_0$, determines the resulting size distribution of the helium clusters and the average flux. 

Atoms or molecules can be incorporated in the droplets by pickup from a gas cell and by using this technique, small molecules \cite{Vilesov_OCS} and biomolecules \cite{Lindinger:1999vn, Scheier_nucleobase} as well as large species such as C$_{60}$  \cite{Federmann_C60} have successfully been doped into helium droplets. Using laser vaporization, less volatile materials such as refractory metal atoms can be evaporated and clusters can grow inside helium droplets \cite{Stienkemeier_Cluster, Doppner_Rostock_Metal_He}. Those clusters, however, will occur in a distribution of sizes that is governed by Poisson statistics.

For studying molecules in helium droplets, a prerequisite is the ability to bring the intact molecule into the gas phase. For many interesting species such as most larger biomolecules, this can not be done via evaporation. Pulsed laser desorption is one other possibility, doing so, however, yields low concentrations in the gas phase as well as frequently a mixture of species that additionally contains decomposed molecules and matrix molecules. More promising would be to use established techniques, as for example electrospray ionization followed by mass separation, and to incorporate those mass/charge selected ions into helium droplets. 

We here present an experimental approach in which mass/charge selected ions that are stored and accumulated in an ion trap are picked up by helium droplets traversing the trap. The approach is conceptually similar to pickup experiments of neutrals from gas cells, however a crucial difference is that in the case of the ion trap, use is made of the high kinetic energy of the heavy helium droplets, which allows ions only to leave the trap once they are inside a droplet. 
The so doped droplets can then be further investigated and here, an initial study on the droplet masses and beam velocities is presented.
Having biomolecules embedded in helium droplets will allow many exciting experiments, such as spectroscopic investigations and possibly in the future single molecule diffraction experiments using x-ray free electron lasers \cite{Hajdu_xray_short, Chapman, Rost_NeonCluster} or electron diffraction experiments on aligned biomolecules \cite{Spence:2004kx}. For large molecules inside helium droplets, recent spectroscopic studies have shown to give spectra of much higher resolution, compared to those obtained for the bare gas-phase molecule in a molecular beam setup \cite{Drabbels_FEL}.

\begin{figure}
\centering \includegraphics[width=1.0\columnwidth]{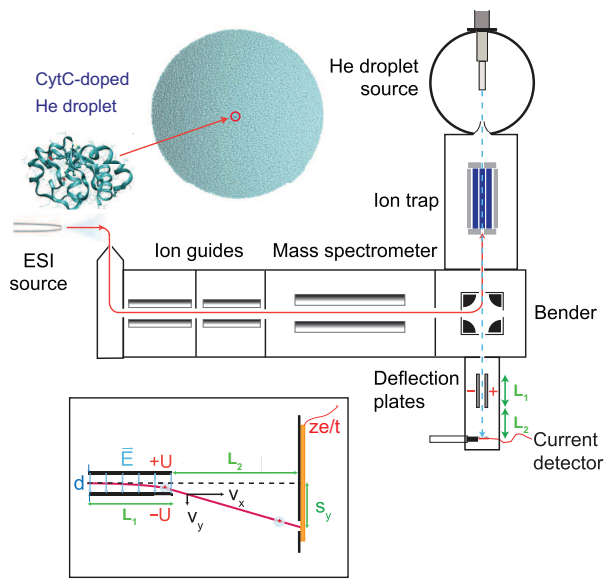}
\caption{Scheme of the experimental setup. Ions are brought into the gas phase via electrospray ionization, are mass/charge selected by a quadrupole MS and stored in an ion trap. Helium droplets can pick up those ions and after some distance downstream, the direct ion current is measured. Electric field deflection can be used to determine the droplet size distribution (see text). 
}
\label{fig1}
\end{figure}

A scheme of the experimental setup is shown in fig.\ref{fig1}. Ions are brought into the gas phase via electrospray ionization (ESI) and mass/charge selected in a quadrupole mass spectrometer. This part of the machine is a modified version of a QTof Ultima (Waters Corporation). The ion beam can then be bent 90$^\circ$ by a quadrupole bender to be injected into a linear hexapole ion trap. Alternatively, when the bender is turned off, the ions are transferred into a time of flight mass spectrometer (not shown in fig.\ref{fig1}). The ion trap consists of six 30 cm long, 5 mm diameter rods which are placed on a 14.1 mm diameter circle. The trap is operated with 200 $V_{p-p}$, 1.7 MHz RF and two trapping electrodes that are kept 1-- 3 volts above the rod DC level are used for longitudinal trapping.

The helium droplet source is pulsed and similar in design to that of A. Vilesov \cite{Vilesov_pulsed}. In brief, a commercial pulsed valve (Parker Series 99) with a Kel-F poppet is mounted to the low temperature stage of a closed-cycle He cryocooler (Sumitomo RDK 408D2). To improve the cooling of the nozzle, its faceplate is machined from copper with an inserted stainless-steel piece which provides the sealing to the main valve body. The valve opening is a 0.8 mm diameter hole followed by a 90$^\circ$, 1 cm long cone. 
The valve temperature can be adjusted between 4.7 K and 30 K, the helium pressure can be regulated up to 50 bar and the valve is typically operated with $\approx$ 200 $\mu$s long pulses at a 4 Hz repetition rate.
Five cm downstream the nozzle, the beam is skimmed (Beam dynamics model 2, 2 mm opening) and enters the ion trap. The droplets move slowly ($\approx$200 -- 300 ms$^{-1}$), however their mass can be large and so will be their kinetic energy. When a droplet picks up an ion, the kinetic energy of the doped droplet will exceed the longitudinal trapping potential energy and, inside the droplets, the ions can leave the trap.
The ion-doped helium droplets pass straight through the bender and can be analyzed and detected further downstream. The charged droplets can be deflected by applying voltages to a pair of parallel metal plates, and the ion current can be measured 50 cm after the plates on a movable copper plate. Due to the high droplet mass in the here presented experiments, it is not possible to use sensitive charged particle detectors that rely on electron release from a surface upon ion impact as the first step. Instead, the direct ion current is measured via a current to voltage amplifier. Two types of amplifiers are used. One (FEMTO DLPCA-200) has a calibrated maximal gain of $10^{11}$ V/A and a bandwidth of 1.1 kHz. The other one (AMPTEK A250) provides a better signal to noise, however is not calibrated and is the one used in most of the here presented experiments.

In a typical experiment, a pulse of He buffer gas is injected into the ion trap. Incoming ions with a kinetic energy just barely above the trapping potential energy loose energy via collisions and are stored in the trap. After a few seconds, depending on the ion current, the number of ions in the trap increases no further as presumably the space-charge limit ($10^6 - 10^7 e\cdot cm^{-3}$) is reached. Then, the bender is turned off, helium droplets are generated for 15 seconds and the ion current of charged droplets is measured as a function of time after valve opening. 
After these 15 seconds, the number of ions in the trap is reduced by the helium droplets to about 50\% of its initial value (measured in the time of flight mass spectrometer) \cite{MS_suppl} and the cycle is repeated after reloading the trap. At the ion densities in the trap, the pickup of multiple ions is unlikely.\\
The experiments are performed on droplets doped with protonated singly charged Phenylalanine (Phe) as well as with the 104 amino acid protein Cytochrome C (CytC) in different charge states. This protein  (Aldrich, horse heart) has a mass of 12327 amu and is known to be present in a broad range of charge states after electrospray, ranging from z=+9 up to z=+17. In the here presented experiments, the charge states z=+9, +14 and +17 are selected. 

\begin{figure}
\centering \includegraphics[width=1\columnwidth]{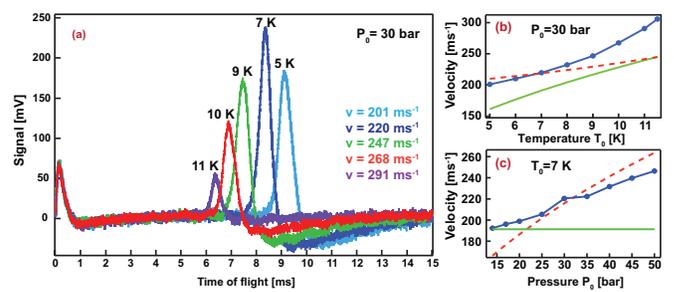}
\caption{(a) Time of flight profiles of CytC ions (z=+14) embedded in He droplets. The He droplets are produced at P$_0$= 30 bar and varying source temperatures T$_0$ from 5 K to 11 K. With increasing T$_0$ the peak shifts to earlier arrival times. A considerable portion of the peak widths can be attributed to the time response of the current amplifier.\\
Measured velocities (blue points and curve) of ion-doped helium droplets as a function of source temperature T$_0$ (b) and pressure P$_0$ (c). Also shown are predictions for an ideal He gas expansion (green solid curve) as well as a liquid expansion (red dashed curve) following the Bernoulli expression.}
\label{fig2}
\end{figure}
In fig.\ref{fig2}(a), time of flight profiles of helium droplets doped with CytC in the charge state +14 are shown as a function of source temperature $T_0$.
At early times, an interference caused by the trigger pulse of the valve can be observed. The ion signal appears at around 8 ms and has a width of $\approx$1 ms. Measuring the current with the calibrated amplifier gives peak currents of up to $\approx$ 20 pA, which implies about $10^4$ ion-doped droplets per pulse. 
Such intensities should be sufficient to investigate the dopant ions via action spectroscopy (in which the wavelength dependent response of the chromophore upon light exposure is monitored) for which only few ions per pulse are needed  \cite{Duncan:2000uq}.
All traces in fig.\ref{fig2} (a), are taken at P$_0$= 30 bar. The lowest temperature investigated is 5 K. When increasing the temperature, the peaks shift to earlier times. At temperatures above 10 K, the intensity rapidly diminishes and no doped droplets can be observed above 11.5~K. The measured beam velocities as a function of source temperature $T_0$ are shown in fig.\ref{fig2}(b) and range from 200 ms$^{-1}$ at 5 K to 305 ms$^{-1}$ at T$_0$ =11.5~K. 
The droplet velocities are also observed to vary with source pressure $P_0$ (see fig.\ref{fig2}(c)). At a fixed temperature of 7~K, the velocity increases from 190 ms$^{-1}$ at 20 bar to 240 ms$^{-1}$ at 50 bar and below 20 bar, no signal is observed. For comparison, the expected velocities for a supersonic expansion of an ideal gas and the expected velocities from the Bernoulli equation $v=c \sqrt{2P_0/\rho _0}$, with the discharge coefficient c as 1.0 and with $\rho _0$ as the known values for the helium densities, are shown. Clearly, the ideal gas expression does not predict the observed pressure dependence, which is qualitatively reproduced by the Bernoulli expression. However, the ideal gas expression does better in predicting the temperature dependence of the beam velocities.

Important parameters are the size and size distribution of the ion-doped droplets. For an experimental determination, a pair of oppositely charged metal plates with a spacing of d=7 mm and a length of $L_1$=10 cm is installed behind the trap and the bender (see fig.\ref{fig1}), about 120 cm downstream of the droplet source. Just in front of the plates, a 1.2 mm collimating slit is located.
Applying a voltage $\pm U$ to the plates causes a deflection $s_y$ of the beam which depends on beam velocity $v_x$, droplet mass $m$ and droplet charge $z e$ according to:
\begin{equation}
s_y = \frac{z e E L_1(L_1+2L_2)}{2  m  v_x^2}
\label{eq1}
\end{equation}
where $E=2U/d$.
The deflection is detected by measuring the ion current on a plate after a movable 1 mm slit at a distance $L_2$=50 cm from the end of the deflection plates. The droplet source temperature and pressure are kept at 8 K and 30 bar, respectively. 

\begin{figure}
\centering \includegraphics[width=1\columnwidth]{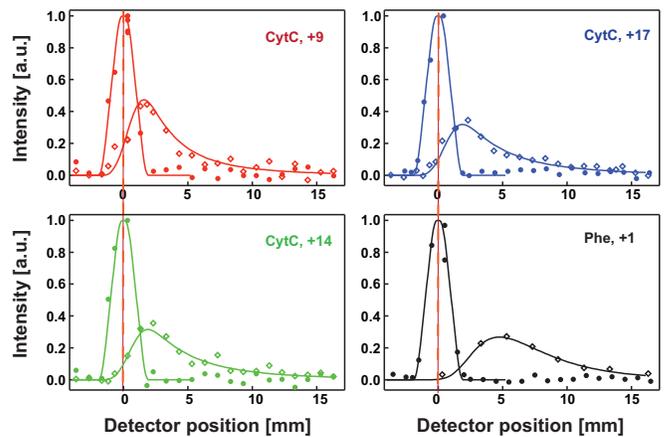}
\caption{Deflection profiles for three different charge states of CytC z=+9, +14 and +17 and protonated singly charged Phe at an electric field of E=5.7 kV/cm (diamonds) and at zero field (filled circles). Also shown are the respective fitted curves (lines). 
}
\label{fig3}
\end{figure}

To measure the deflection for a particular electric field E, time of flight profiles are recorded as a function of detector position. In fig.\ref{fig3}, the integrated ion signal for the three charge states of CytC as well as protonated Phe at a field of 5.7 kV/cm (diamonds) and with no deflection field applied (filled circles) are shown.
It is apparent that applying an electric field causes the droplet beam profiles to shift and to broaden. At a constant field, the observed deflection should depend on the droplet mass and charge and higher charged droplets should give rise to a larger deflection. Strikingly, when comparing the highly charged CytC doped droplets to the singly charged Phe doped droplets, the opposite is observed. Comparing the differently charged CytC doped droplets among each other, an increased deflection with increasing charge is observed. The increased deflection of the Phe doped droplets must thus result from a smaller droplet size.\\
In order to quantify the observations, the curves in fig.\ref{fig3} are fitted with a model. The curves with no field applied are fit with an instrumental function that contains the widths of the two slits as well as the absolute zero position. The model for the fits of the deflection curves contains the same instrumental function as well as a size distribution function of the droplets. For that, we take a log-normal distribution, which has previously been shown to be applicable for helium droplets \cite{Schilling}. Such a distribution is described by a mean size as well as by a width parameter. The width parameter $\sigma$, the standard deviation of the distribution of the logarithm of the droplet size, is kept constant at 1.0 for all CytC curves and at 0.58 for the Phe curve. The resulting (arithmetic) mean droplet masses are ($2.3, 2.7$ and $3.2) \cdot 10^{12}$ amu for CytC +9, +14 and +17, respectively and $2.2\cdot 10^{10}$ amu for Phe doped droplets. 

It should be noted that in most He-droplet experiments reported to date, it is notoriously difficult to directly measure the droplet size distributions. Nonetheless, the here observed droplet sizes are considerably larger than those observed in this temperature and pressure range by others \cite{Buchenau_expansionRegimes, Northby_supercriticalFragm, Schilling, Slenczka_UziEven} and more similar to those obtained at much lower temperature from the breakup of a liquid helium beam \cite{Toennies_ultralarge}.
The observation that the here observed droplet sizes are orders of magnitude larger could be explained by the larger nozzle orifice of 800 $\mu m$, compared to 500 $\mu m$ used in the experiments of Vilesov \cite{Vilesov_pulsed}. In addition, the here employed nozzle is machined out of copper (compared to stainless steel used in other experiments \cite{Vilesov_pulsed}) and has a longer 90 degree conical section, which might further promote the formation of large droplets. 

An interesting and puzzling observation is the difference in droplet sizes in the CytC and Phe experiment. As the droplet size distribution emitted from the source is the same in both experiments, the pick up efficiency for a given droplet size must depend on the dopant molecule mass and/or charge. Apparently, a droplet size of $\approx 10^{12}$ atoms is required to pick up CytC ions and the curves in fig.\ref{fig3} thus represent only the tail of the real size distribution. Whether the curve for Phe represents the droplet size distribution as emitted from the source is not yet clear. \\
The question is which physical factors could give rise to such a size effect. The velocity of CytC in the trap can be neglected relative to the velocity of the helium droplets; the protein thus dives into liquid helium with about the droplet velocity (200--300 $ms^{-1}$). A rough estimation based on an impulsive model gives that in a distance of about 0.02 $\mu m$, the particle is slowed down to the Landau velocity (60 $ms^{-1}$), after which it could move without much friction through the droplet. When having a droplet of 4 $\mu m$ diameter ($\approx 10^{12}$ atoms), the protein reaches in $\approx 70~ns$ the other edge of the droplet. 
Before the pickup by helium droplets, the proteins are thermalized in the ion trap at 300 K. Their internal vibrational energy is then $\sim$ 30 eV, estimated using typical values for the heat capacity of a protein \cite{Protein_Cv}. After some time, this energy will be transferred to the helium, causing about $4\cdot 10^{4}$ helium atoms to evaporate \cite{Rev_Angewandte_Toennies_Vilesov}. Recent observations have shown that vibrationally excited small ions can leave helium droplets, presumably assisted by a shell of "hot" helium surrounding them \cite{Drabbels_ionEject}. 
A possibility is thus, that times significantly smaller than $70~ns$ are insufficient to cool the protein and its surroundings, and when the still "warm" protein reaches the edge of the droplet, it will escape again. It should be noted that in the here presented CytC experiments, the internal energies deposited at once by a dopant molecule into a helium droplet are considerably larger than those in all previous helium droplet experiments.
Protonated Phe has a much lower internal vibrational energy than that of CytC and therefore does not require long cooling times or large droplets to become trapped.

In summary, we have presented a new technique that allows for the first time the pickup of large mass/charge selected biomolecular ions by helium droplets. The droplet sizes are determined via electrostatic deflection and are found to be very large, containing more than 10$^{10}$ helium atoms and being larger than 1 $\mu m$ in size. The approach is general and allows the selective and clean incorporation of a wide range of other mass/charge selected species, such as for example cluster ions, into helium droplets and thus opens the helium droplet isolation technique to a wide range of new species. 

We gratefully thank Andrey Vilesov for help and advice with the pulsed droplet source and Waters for providing us the QTof mass spectrometer.
\bibliographystyle{prsty}

\end{document}